\documentclass[fleqn,usenatbib]{mnras}
\usepackage[T1]{fontenc}
\usepackage{ae,aecompl}
\usepackage{graphicx}	% Including figure files
\usepackage{amsmath}	% Advanced maths commands
\usepackage{amssymb}	% Extra maths symbols
\title[Peculiar motion from Hubble diagram of quasars]{Solar system peculiar motion from the Hubble diagram of quasars and testing the Cosmological Principle}
\author[A. K. Singal]{Ashok K. Singal\thanks{E-mail: ashokkumar.singal@gmail.com}\\
{Astronomy and Astrophysics Division, Physical Research Laboratory, 
Navrangpura, Ahmedabad - 380009, India}}
\date{Accepted XXX. Received YYY; in original form ZZZ}
\pubyear{2021}
\begin{document}
\label{firstpage}
\pagerange{\pageref{firstpage}--\pageref{lastpage}}
\maketitle

% Abstract of the paper
\begin{abstract}
We determine here peculiar motion of the Solar system, first time from the $m-z$ Hubble diagram of quasars.
Observer's peculiar motion causes a systematic shift in the $m-z$ plane between sources lying along the velocity vector and those in the opposite direction, providing a measure of the peculiar velocity. Accordingly, from a sample of $\sim 1.2 \times 10^5$ mid-infrared quasars with measured spectroscopic redshifts, 
we arrive at a peculiar velocity $\sim 22$ times larger than that from the CMBR dipole, but direction matching within $\sim 2\sigma$. 
Previous findings from number count, sky brightness or redshift dipoles observed in samples of distant AGNs or SNe Ia too had yielded values two to ten times larger than the CMBR value, 
%but this by far is the largest value arrived at for the peculiar motion, 
though the direction in all cases agreed with the CMBR dipole.
Since a genuine solar peculiar velocity cannot vary from one dataset to another, an order of magnitude, statistically significant, discordant dipoles, might imply that we may instead have to look for some other cause for the genesis of these dipoles, including that of the CMBR. 
At the same time, a common direction for all these dipoles, determined from completely independent surveys by different groups employing different techniques, might indicate that these dipoles are not resulting from some systematics in the observations or in the data analysis, but could instead suggest a preferred direction in the Universe due to an inherent anisotropy, which, in turn, would be against the Cosmological Principle (CP), the most basic tenet of the modern cosmology.
\end{abstract}
%--------------------------
\section{INTRODUCTION}
According to the CP, the universe should appear isotropic, without any preferred directions, to a comoving observer, having no peculiar motion relative to the cosmic fluid of the expanding universe. 
To such an observer the average properties of a given class of distant astronomical objects in the universe should appear statistically to have similar distributions in various directions.
 However a peculiar motion of such an observer might introduce a dipole anisotropy in the observed properties of a class of objects and which, in turn, might be exploited to infer the peculiar velocity of the observer. 
For instance, the Cosmic Microwave Background Radiation (CMBR) shows an isotropic distribution except for a dipole anisotropy which, when ascribed to a peculiar motion of the observer, has given a peculiar velocity of the solar system to be $370$ km s$^{-1}$ along RA$=168^{\circ}$, Dec$=-7^{\circ}$ in the sky (Lineweaver et al. 1996; Hinshaw et al. 2009; Aghanim et al. 2018). 
%\cite{1,2,3}.

Another such quantity that could be employed to look for departures from isotropy of the universe is the angular distribution of distant radio sources in the sky. This could provide an independent 
check on the interpretation of CMBR dipole anisotropy being due to the observer's motion since an effect of the observer's motion should show up as a dipole anisotropy in sky distribution of the radio source population too.
Also, while CMBR provides information 
about the isotropy of the universe for redshift $z \sim 1100$, the radio source population refers to a 
much later epoch $z \sim 1-3$. Thus it also provides an independent check on the CP for the matter universe, as isotropy of the Universe is assumed for matter and radiation {\em for all epochs}.
In last one decade, peculiar motions determined from the number counts, sky brightness or redshift distributions in large samples of distant active galactic nuclei (AGNs) have yielded peculiar velocities two to ten larger than that from the CMBR, although in the same direction as the CMBR dipole (Singal 2011; Rubart \& Schwarz 2013; Tiwari et al. 2015; Colin et al. 2017; Bengaly, Maartens \& Santos 2018; Singal 2019a,b; Secrest et al. 2021; Siewert, Rubart \& Schwarz 2021; Singal 2021a,b).
%\cite{4,5,6,7,8,9,10,11,Si21a,Si21b}. 
A more recent determination of the peculiar velocity of the Solar system from a sample of supernovae type Ia (SNe Ia), has yielded a value about 4 times the CMBR value, again along the same direction (Singal 2021c).

These findings of the AGNs and SNe Ia dipoles being much larger than the CMBR value (Singal 2011; Rubart \& Schwarz 2013; Tiwari et al. 2015; Colin et al. 2017; Bengaly et al. 2018; Singal 2019a,b; Secrest et al. 2021; Siewert et al. 2021; Singal 2021a,b,c) 
%\cite{4,5,6,7,8,9,10,11,Si21a,Si21b,Si21c}, 
cast doubt on these dipoles, including the CMBR, being the ultimate representatives of the solar peculiar motion. %While the CMBR dipole refers to the radiation era ($z \sim 1100$), AGNs and SNe Ia  dipoles represent the much later matter era ($z \sim 1-3$). 
The matching of their directions, however, shows that different dipoles are not because of some totally independent, random origins, nor could we say that the CMBR dipole is the one which is to be considered as more fundamental than the other ones for establishing the CP.
Because of the impact of any genuine variation in the peculiar motion, determined from different samples or techniques, could be of such wide ramifications, it is important to be able to get peculiar motion from independent data, and possibly using independent techniques.

Here, we determine peculiar motion of the Solar system, first time from the magnitude-redshift ($m-z$) Hubble diagram of quasars.
%, drawn originally from the CATAWISE survey of more than a billion objects.  
Owing to the Doppler effect, arising from the peculiar motion of the observer, the observed intensity and redshift of a distant quasar get modified, giving rise to a small but finite displacement of a source in the $m-z$ plane. The extent and direction of the displacement in the $m-z$ plane is determined by the projection of the observer's peculiar motion along the sky position of the quasar. The quasars lying in the hemisphere centred on the pole of the peculiar motion undergo in the $m-z$ plot displacements opposite to those of the quasars lying in the opposite hemisphere, centred on the anti-pole. 
As a result, in the Hubble diagram, there occurs a systematic shift between sources lying in the hemisphere along the peculiar motion and those lying in the opposite hemisphere and this shift could provide a measure of the peculiar velocity of the observer.  Such a technique was applied recently (Singal 2021c) to a sample of supernovae type Ia (SNe Ia), to determine the peculiar velocity of the Solar system from the magnitude-redshift Hubble diagram for the SNe~Ia, one of the best standard candles known. Here we apply this technique to determine the peculiar velocity of the Solar system, first time from the magnitude-redshift Hubble diagram of quasars. As is well known, unlike SNe Ia, there is a large spread in the  Hubble diagram of quasars, therefore we would require a much larger sample to get any meaningful values for the peculiar motion. Moreover it would be helpful if the sample of quasars has the magnitude determination from a single instrument to minimize the spread in magnitude values  from errors due to differential calibrations and any systematic differences in different instruments at different colors.
Accordingly, we determine our peculiar motion from a large sample of $\sim 1.2 \times 10^5$ quasars, with mid-infrared magnitudes and measured spectroscopic redshifts.

%In last one decade, peculiar motions determined from the number counts, sky brightness or redshift dipoles  observed in large samples of distant active galactic nuclei (AGNs) (Singal 2011; Rubart \& Schwarz 2013; Tiwari et al. 2015; Colin et al. 2017; Bengaly, Maartens \& Santos 2018; Singal 2019a,b; Secrest et al. 2021; Siewert et al. 2021; Singal 2021a,b) 
%\cite{4,5,6,7,8,9,10,11,Si21a,Si21b} 
%have yielded peculiar velocities  two to ten larger than 370  km s$^{-1}$, the value determined from the dipole anisotropy in the Cosmic Microwave Background Radiation (CMBR) \cite{1,2,3}, though in all cases the directions inferred matched within statistical uncertainties with the CMBR dipole along  RA$=168^{\circ}$, Dec $=-7^{\circ}$.  
%-------------------------------------------------------
\begin{figure*}
\includegraphics[width=\linewidth]{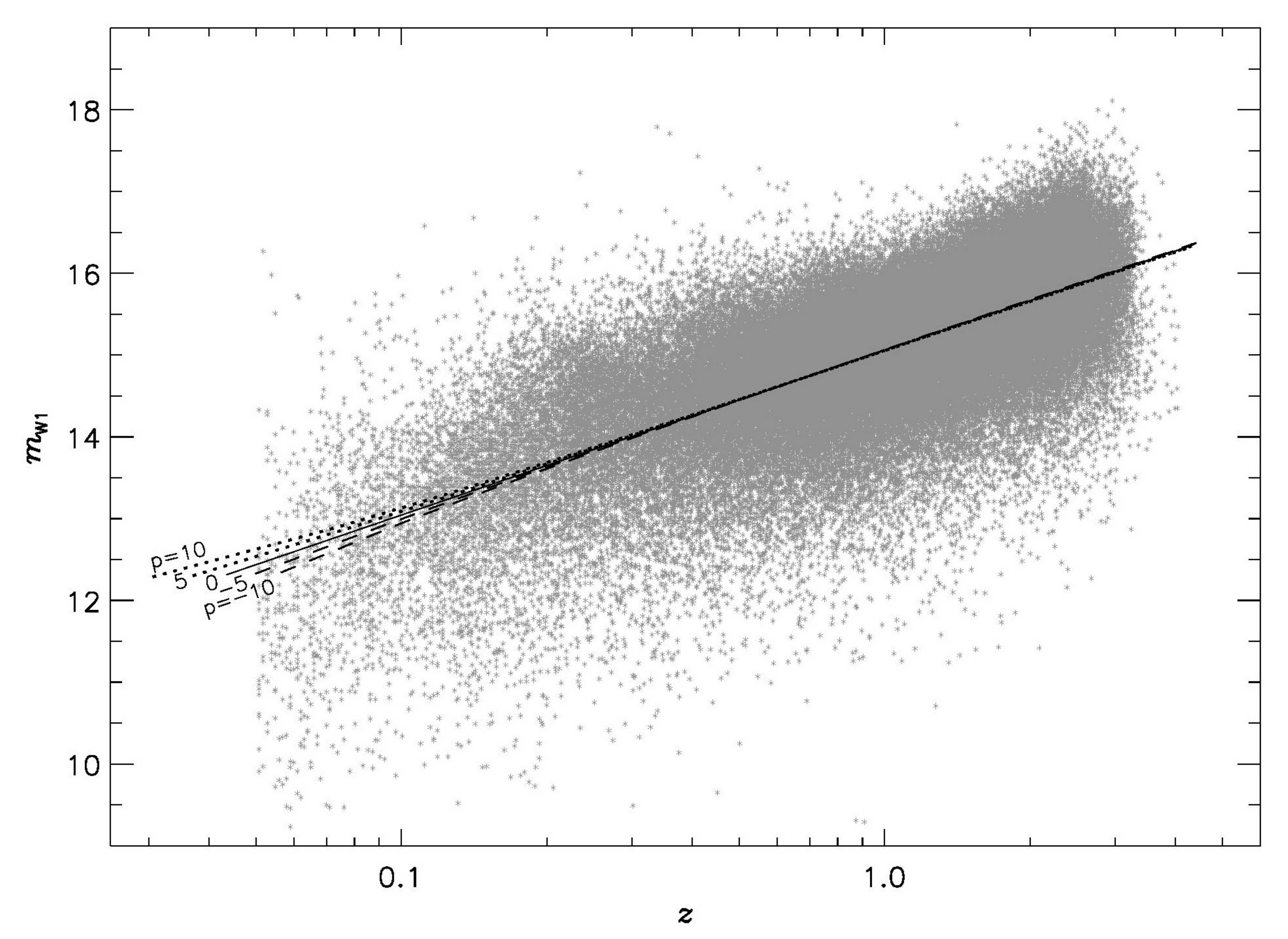}
\caption{The observed Magnitude-redshift Hubble ($m_{\rm w1}-z$) diagram for our sample of quasars with known spectroscopic redshifts.
The continuous line in the middle shows the best fit ($m_{\rm w1} \propto \log z$) to all quasars in our sample. Due to observer's  peculiar velocity, assumedly along the CMBR dipole, individual sources at any point in the $m_{\rm w1}-z$ diagram would get displaced, with the displacement being, to a first order, directly proportional to the amplitude of the peculiar velocity, assumed to be a small non-relativistic value. 
To get an idea of the loci of the expected displacements, the dotted lines above the continuous line (i.e., at higher $m_{\rm w1}$) show the displacements expected for different amplitudes of the peculiar velocity (quantified by $p$, in units of the CMBR value of 370 km s$^{-1}$), for sources lying in the direction (pole) of the  peculiar velocity at its apex, while the dashed lines below the continuous line show loci of the displacements expected for sources lying in the anti-pole direction, for different $p$ values ($p<0$ in the anti-pole direction). It is clear that the displacements become appreciable only at low redshifts ($z \stackrel{<}{_{\sim}}  0.5$).}
\end{figure*}
%--------------------------------------------
%%-------------------------------------------------------
\begin{figure*}
\includegraphics[width=\linewidth]{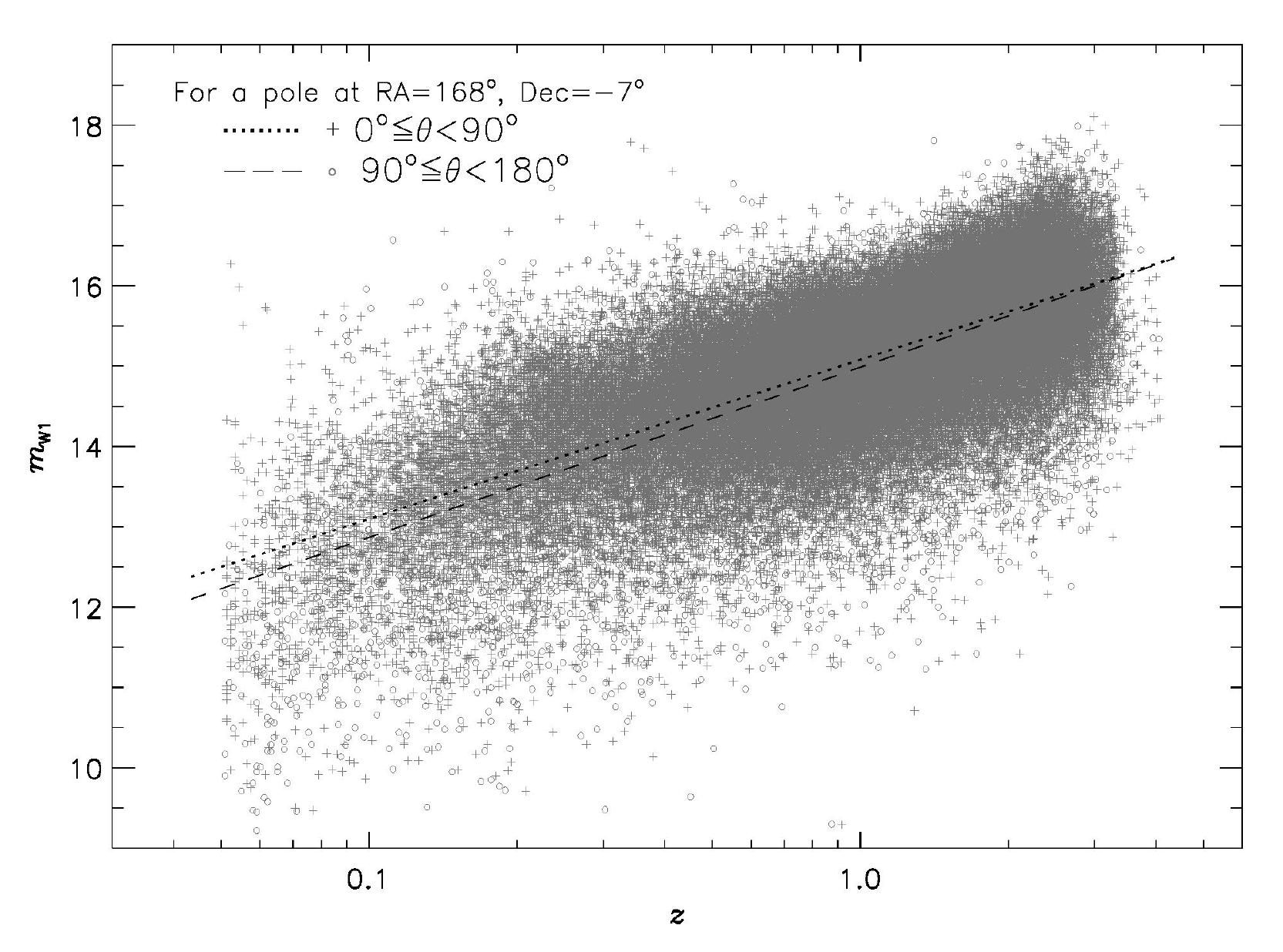}
\caption{The observed Magnitude-redshift Hubble ($m_{\rm w1}-z$) diagram  for our sample of quasars with known spectroscopic redshifts.
The dotted line shows the straight line ($m_{\rm w1} \propto \log z$) fit to quasars (`+') lying within the hemisphere $\Sigma_1$, centered on the CMBR dipole direction, while the dashed line is for the quasars (`o') in the opposite hemisphere, $\Sigma_2$. From the CP, the two datasets are expected to be statistically similar, otherwise in all respects, and the two lines should then be coinciding. The finite displacement between the two straight line-fits, presumably therefore, is due to observer's  peculiar velocity, assumed to be a small non-relativistic value, with the displacement, to a first order, being proportional to the peculiar velocity component along the CMBR dipole.}
\end{figure*}
%----------------------------------------
\begin{figure*}
\includegraphics[width=\linewidth]{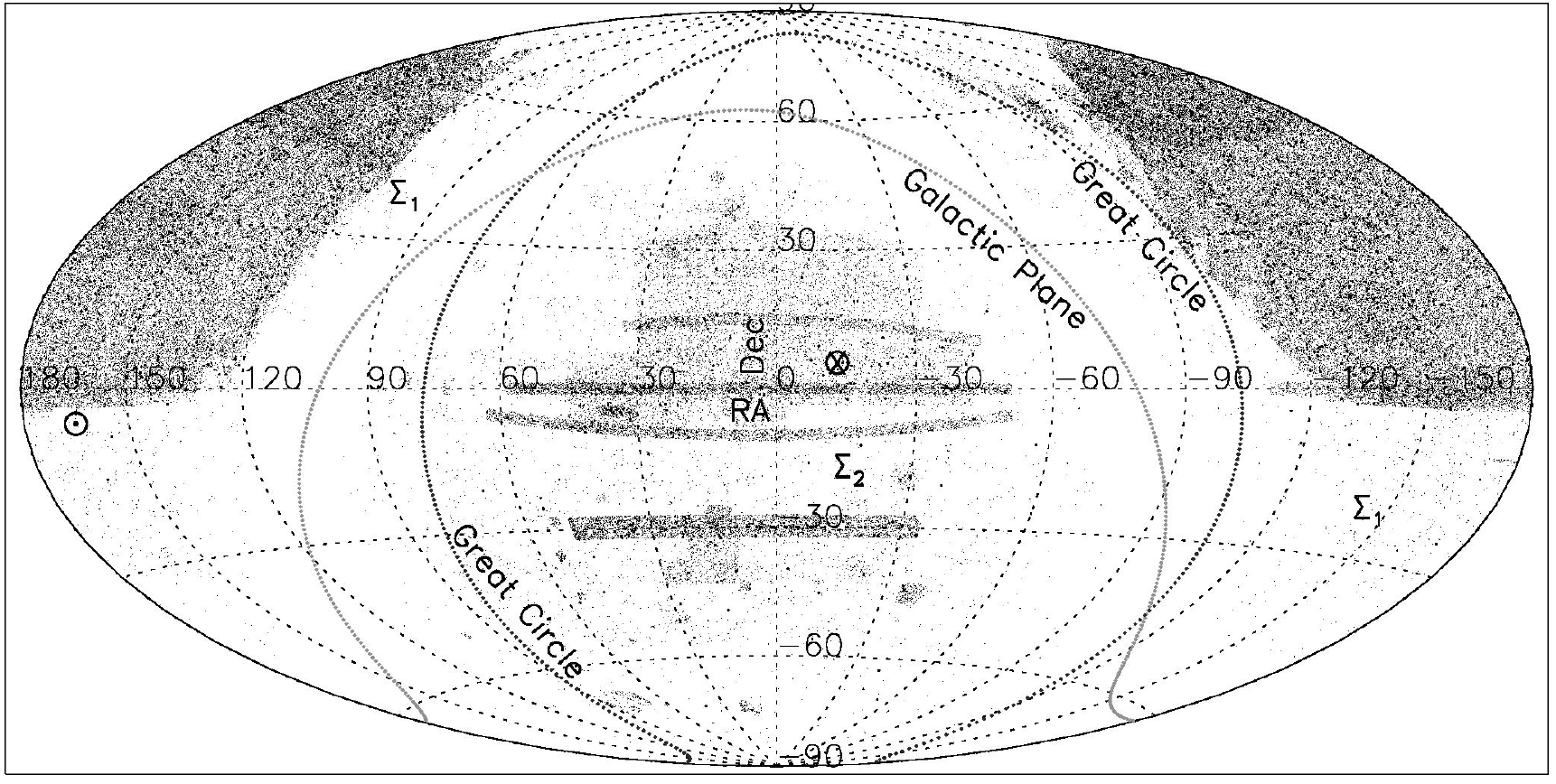}
\caption{The sky distribution of $\sim 1.2 \times 10^5$ quasars in our sample, in~the Hammer--Aitoff equal-area projection map, plotted in equatorial coordinates with right ascension (RA) from $-180^\circ$ to $180^\circ$ and declination (Dec) from $-90^\circ$ to $90^\circ$. The distribution, as determined by the selection criteria for the spectroscopic redshift determination in various sub-samples, is not uniform across the sky. The~sky position of the CMBR pole is indicated by $\odot$, while the corresponding anti-pole is indicated by $\otimes$. The great circle dividing the two hemispheres $\Sigma_1$ and $\Sigma_2$ is shown. Also shown is the galactic plane.
}
\end{figure*}
%----------------------------------------------------------------

%--------------------------------------------
\section{Peculiar motion from the Hubble diagram of quasars}
According to the CP, an observer comoving with the cosmic fluid would find observed relationship  between any  properties of distant objects, e.g., the redshift-magnitude Hubble relation for quasars, apart from any statistical scatter, to be independent of direction. 
On the contrary, an observer moving with a peculiar velocity, might find  a dipole anisotropy in some of the observed properties. 
By observing such variation over sky for a sufficiently large sample of sources, one could compute the peculiar velocity of the observer with respect to the comoving coordinates.

Peculiar velocity of the observer gives rise to the Doppler effect, modifying the observed redshift and optical magnitude of an object.  We assume the peculiar velocity $v$ of the observer to be non-relativistic ($v\ll c$), as is indicated by all previous measurements (Aghanim et al. 2018; Singal 2011,19a,b; Secrest et al. 2021; Singal 2021b,c),
%\cite{3,4,9,10,11,Si21b,Si21c},
$c$ being the speed of light in vacuum. 
For a source lying at an angle $\theta$ with respect to the direction (pole) of the peculiar motion, as seen by the observer,
the Doppler factor $[\gamma(1-(v/c)\cos\theta]^{-1}$, with $\gamma=\surd(1- (v/c)^2)$ being the Lorentz factor, reduces for $v\ll c$ to $[(1-(v/c)\cos\theta]^{-1}$.  
Then the observed flux density of the source would be affected by 
\begin{eqnarray}
\label{eq:80.3}
S=S_{\rm o}(1-(v/c)\cos\theta)^{-2}, 
\end{eqnarray}
where $S_{\rm o}$ is the  flux density as measured by a comoving observer, i.e., for a nil peculiar motion of the observer. 
Here one factor of $(1-(v/c)\cos\theta)$ arises because of the shift in the frequency of each photon, while another similar factor arises from the change in number of photons arriving per unit time at the observer.
It should be noted that one expects the ratio between the number densities  as well as the sky intensities for the observer with a peculiar motion and the one stationary with respect to the comoving coordinates to have another factor of $(1-(v/c)\cos\theta)^{-2}$  because of aberration, due to which the observed differential solid angle will be narrower by a factor
\begin{eqnarray}
\label{eq:80.4}
{\rm d}\Omega={\rm d}\Omega_{\rm o} (1-(v/c)\cos\theta)^2\;. 
\end{eqnarray}
where ${\rm d}\Omega_{\rm o}$ is the solid angle for the stationary observer. Therefore, an observer with a peculiar motion, in comparison to a comoving observer at the same location in the cosmic fluid, would observe higher number densities as well as sky intensities. However, the observed flux density of individual sources will not be affected by any additional factor due to aberration. 

Then the observed redshift and optical magnitude of an individual source are given by (Davies et al. 2011)
\begin{eqnarray}
\label{eq:80.1}
(1+z)=(1+z_{\rm o})(1-v\cos\theta/c), \\
\label{eq:80.2}
m=m_{\rm o}+5\log(1-v\cos\theta/c), 
\end{eqnarray}
where $m_{\rm o}$ and $z_{\rm o}$ are the values as would be measured by the comoving observer, without a peculiar motion.  From the CP, $m_{\rm o}$ and $z_{\rm o}$ should have isotropic distributions.

%-------------------------------------------------------

For a given $v$ of the observer, different sources, depending upon their $\theta$, will get displaced in the $m-z$ plot differently. As the effects on both $m$ and $z$ are proportional to $\cos\theta$, all source with $\cos\theta>0$, and thus lying in a hemisphere, say $\Sigma_1$, centred on the pole, will get displaced in the $m-z$ plot  opposite to the sources with $\cos\theta<0$ and thus lying in the opposite hemisphere, say $\Sigma_2$ centred on the anti-pole. Accordingly, there will be a systematic displacement between sources belonging to $\Sigma_1$ and $\Sigma_2$, and we could utilize this shift to get a handle on the peculiar velocity of the observer.
%--------------------------------------------
\section{Our sample of quasars}
Our sample of quasars is selected from a larger all-sky sample of 1.4 million active galactic nuclei (AGNs), which is publicly available (Secrest et al. 2015), and is derived in turn from the Wide-field Infrared Survey Explorer final catalog release (AllWISE), incorporating data from the WISE Full Cryogenic, 3-Band Cryo and~NEOWISE Post-Cryo survey~(Wright et al. 2010; Mainzer~et~al. 2014). The~WISE survey is an all-sky mid-infrared survey at 3.4, 4.6, 12 and~22 $\mu$m (w1, w2, w3 and~w4) with angular resolutions 6.1, 6.4, 6.5 and 12 arcsec, respectively. We have selected all quasars with measured spectroscopic redshifts from the original sample (Secrest et al. 2015) of $\sim 1.4$ million objects that met a two-color infrared photometric selection criteria for AGNs, out of the AllWISE catalog of almost 748 million objects. 	

The sample of known-spectroscopic redshift quasars, spans a  redshift range 0.01-7.01. However, we have restricted for our purpose the lower limit to 0.05 in order to keep the effect of local bulk flows to a minimum. Further, there is a sharp drop in number of quasars beyond $z>4$, which we have excluded leaving us with a total of 115610 quasars in our sample, all outside the galactic plane ($b>10^\circ$). The infrared magnitude $w1$ provides a uniform measure at a single magnitude band for all quasars in our sample. 

Figure~1 shows a $m_{\rm w1}-z$ plot for all 115610 quasars in our sample, along 
with a best fit of a straight line ($m_{\rm w1} \propto \log z$) to the data. It should be noted that the figure displays the $m_{\rm w1}$ range 9.0-19.0, there are a small number of quasars outside this range, which  are not shown in the figure to avoid getting the displayed magnitude  scale compressed too much, but otherwise all these quasars are used in our computations. In Fig.~1, the continuous line in the middle shows the best fit ($m_{\rm w1} \propto \log z$) to all quasars in our sample. Due to a peculiar motion of the observer, there could be alterations in the $m_{\rm w1}-z$ plot.  

Due to observer's  peculiar velocity, assumedly along the CMBR dipole, individual sources at any point in the $m_{\rm w1}-z$ diagram would get displaced, with the displacement being, to a first order, directly proportional to the amplitude of the peculiar velocity, assumed to be a small non-relativistic value. 
We can use a parameter $p$ to express the peculiar velocity $v$, in units of the CMBR value, so that $v=p \times 370$ km s$^{-1}$, with $p=0$ implying a nil peculiar velocity while $p=1$ implying the CMBR value.
We show the loci of the expected displacements for different $p$ from the continuous line ($p=0$), by dotted lines for sources lying along the direction of the pole and by dashed lines for sources lying along the anti-pole direction.
Figure~1 shows displacements expected for two values of the peculiar velocity ($p=5$ and $p=10$) by two dotted lines, which lie above the continuous line, i.e. at higher $m_{\rm w1}$, and are representatives of the expected displacements for sources lying along the direction (pole) of the  peculiar velocity at its apex. The dashed lines, on the other hand, drawn for two $p$ values, $p=-5$ and $p=-10$, represent loci of the displacements expected for sources lying along the anti-pole direction ($p<0$ in the anti-pole direction) and lie below the continuous line, i.e. at lower $m_{\rm w1}$.
In Fig.~1, plots depicted for various $p$ values are for sources exactly along the pole or anti-pole direction, For an even distribution of sources along various directions within each hemisphere, the net displacement will on average be half of that shown in Fig.~1 for each $p$ value.

Figure 2 shows $m_{\rm w1}-z$ plot for the quasars in our sample, separately for the two hemispheres; a `+' symbol is used for sources lying within the hemisphere  $\Sigma_1$, centered on the pole of the CMBR dipole, while `o' symbol is for sources lying in the opposite hemisphere  $\Sigma_2$, centered on the anti-pole. The dotted line shows the straight line ($m_{\rm w1} \propto \log z$) fit to quasars lying within $\Sigma_1$, while the dashed line is for the quasars lying within $\Sigma_2$. From the CP, the two datasets are expected to be statistically similar  otherwise in all respects, and the two lines should then be coinciding. 
The finite displacement between the two straight line-fits, presumably therefore, is due to observer's  peculiar velocity, assumed to be a small non-relativistic value, with the displacement, to a first order, being proportional to the peculiar velocity component along the CMBR dipole.

If there were no peculiar motion (component) of the observer along that direction, then there should be no systematic shift in the magnitude-redshift diagram between the  two datasets comprising sources belonging to the two opposite hemispheres. 
However due to a  peculiar motion of the observer  there will be deviations in the $m_{\rm w1}-z$ plot, giving rise to the magnitude gap between the two fits. A Comparison of the observed displacement between the dotted and dashed lines with those expected from Fig.~1, assuming the quasars in our sample to be distributed uniformly over all polar angles  with respect to the CMBR pole, suggests a large peculiar velocity $p\stackrel{>}{_{\sim}} 15$, instead of $p=1$, expected from the CMBR dipole value. This inferred $p$ value, at least to a first order, will be proportional to $\cos \psi$, the projection of the assumed dipole, CMBR here, on the direction of the actual dipole, if the two are different.

The sky distribution of quasars in our sample is displayed in Fig.~3, where we have indicated the positions of the CMBR pole ($\odot$), and its anti-pole ($\otimes$). Also shown are the corresponding great circle dividing the two hemispheres $\Sigma_1$ and $\Sigma_2$ as well as the galactic plane. It is clear that in our sample the distribution of sources among $\Sigma_1$ and $\Sigma_2$ hemispheres is not even by any standards. About $80\%$ of the total quasars in our sample (89660 out of 115610) are in the hemispheres $\Sigma_1$, almost all of these in the northern half of $\Sigma_1$ with only a small fraction in the southern part of $\Sigma_1$. The remainder $\sim 20\%$ (25950 quasars) lie in $\Sigma_2$, with a rather patchy distribution (Fig.~3).
What we really require in our method is that the sources in the sample at various redshifts cover most polar angles reasonably well. Figure~4 shows the redshift ($z$) distribution against the polar angle $\theta$, with respect to the CMBR pole, where quasars with $0^\circ \le \theta\le 90^\circ$ belong to the hemisphere $\Sigma_1$ while those with $90^\circ < \theta\le 180^\circ$  belong to $\Sigma_2$.
%-------------------------------------------------------------------
\section{Pros and cons of the technique}
The technique being applied here for estimating the peculiar motion of the observer from the Hubble diagram  requires a precise knowledge of redshifts and may thus appear to be weaker than the traditional methods (Singal 2011; Rubart \& Schwarz 2013; Tiwari et al. 2015; Colin et al. 2017; Bengaly et al. 2018; Singal 2019a,b; Siewert et al. 2021; Secrest et al. 2021; Singal 2021a,b) of deriving the peculiar motion from the dipole asymmetry in the number counts or sky brightness. However, there are certain advantages here, for it provides an alternate route to the determination of peculiar motion, where the other techniques might not work.
%----------------------------------------
\begin{figure}
\includegraphics[width=\columnwidth]{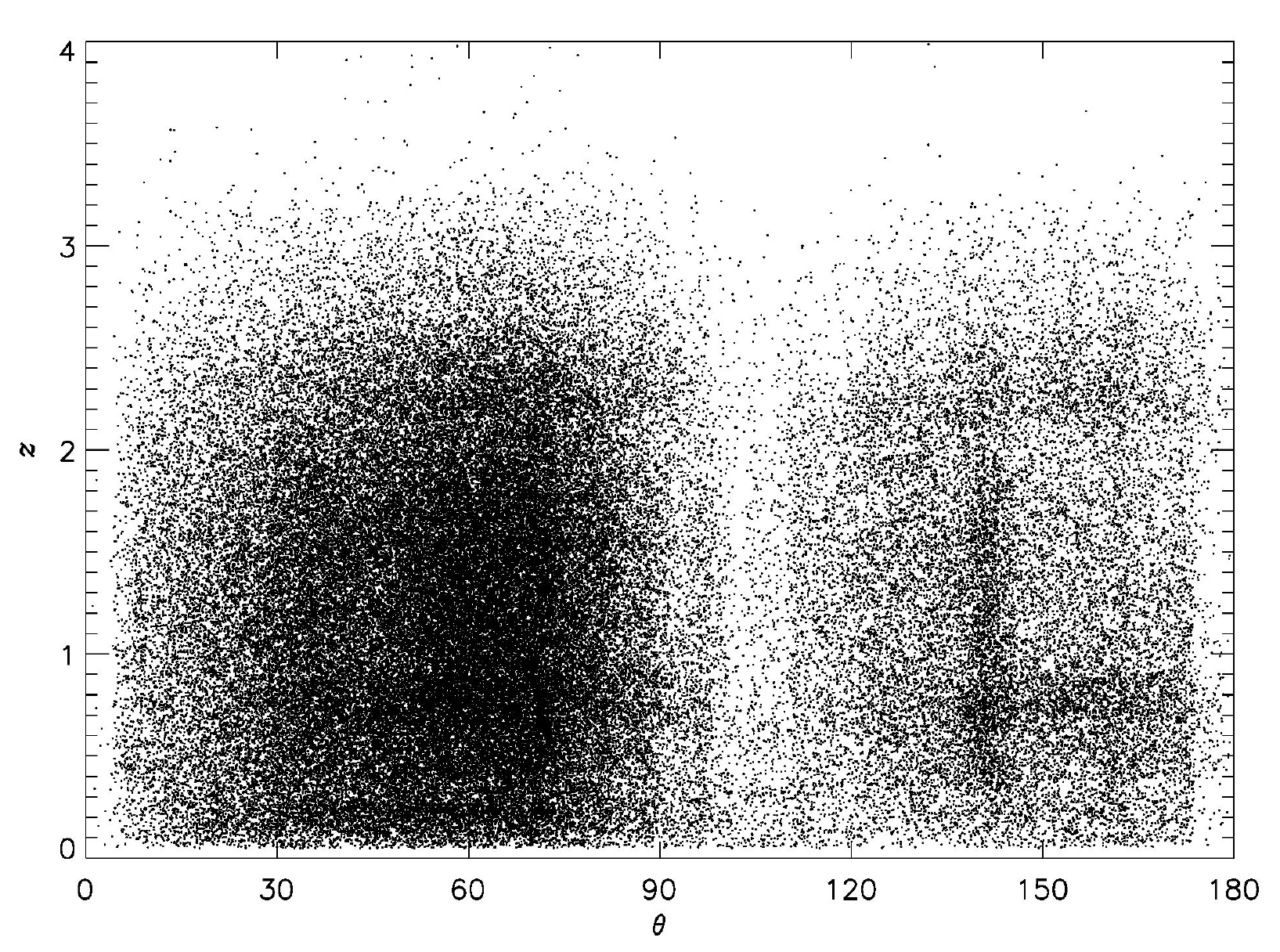}
\caption{Redshift ($z$) distribution of quasars as a function of the polar angle $\theta$ with respect to the CMBR pole. Quasars with $0^\circ \le \theta\le 90^\circ$ belong to the hemisphere $\Sigma_1$ while those with $90^\circ < \theta\le 180^\circ$  belong to $\Sigma_2$.} 
\end{figure}
%-------------------------------------------------------

The latter, more traditional methods, in order to determine the peculiar motion of the observer or equivalently of the Solar system with respect to the average universe, require samples comprising large numbers (millions!) of distant sources. This is because to measure equivalent of the CMBR dipole anisotropy $\sim 10^{-3}$, the number of sources required is of the order of $N\sim 10^6$ to achieve statistical uncertainty $\propto 1/\surd N \sim 10^{-3}$, that would yield a signal to noise of the order of unity (Crawford 2009; Singal 2011,19a). Moreover, a uniform coverage of the whole sky (or at least a predominantly large fraction of it), with possibly only a small number of gaps, made from a single survey (to avoid different calibration scales in different surveys covering different parts of the sky which could in turn affect the number densities of flux-limited samples in different directions differently, especially when the whole exercise depends upon the differential number density of sources observed in different directions to an accuracy better than one part in $10^3$). Of course, a larger sample could help, however, a very careful matching of two or more surveys covering different parts of the sky with different flux-density limits that too at different frequencies of observations with different confusion limits is needed to get consistent results (Colin et al. 2017). Another complication that could arise from a use of two or more different survey data is 
that the dipole amplitudes from individual surveys could be quite different. For instance, the NRAO VLA Sky Survey (NVSS), carried out by the Very Large Array (VLA) of the National Radio Astronomy Observatory (NRAO), comprising a catalog of 1.8 million sources (Condon et al 1998), and  
the TIFR GMRT Sky Survey (TGSS), carried out by the Giant Metrewave Radio Telescope (GMRT, Swarup et al. 1991) of the Tata Institute of Fundamental Research (TIFR), comprising a dataset of 0.62 million sources (Intema et al. 2017), gave dipole magnitudes that differ by more than a factor of two (Singal 2019a). Any overlaps of such samples could yield for the dipole magnitude very different values depending on how much is the overlap in the two such surveys and which survey is covering which particular regions of the sky. Such is seen in Siewert et al. (2021), where the NVSS survey gave $p\sim 4$, while the TGSS survey yielded $p\sim 15$, however, a cross match of both catalogues (TGSSxNVSS) gave $p\sim 10$. 

However the present technique of employing the magnitude-redshift Hubble diagram to estimate the peculiar motion of the observer, does not depend upon a completeness of the survey, nor does it get much affected by a combination of data from a heterogeneous set of various sub-samples. Even a uniform coverage of the whole sky is not a must, the only thing essential is that irrespective of the region of sky a particular source belongs to, the observed properties of individual sources in the sample have not been systematically affected. The conventional wisdom is that the reference frame of the CMBR -- \`a la Cosmological Principle -- is also a reference frame for the average distribution of {\em matter} in the Universe. However, in the last one decade it has been repeatedly seen 
(Singal 2011; Rubart \& Schwarz 2013; Tiwari et al. 2015; Colin et al. 2017; Bengaly et al. 2018; Singal 2019a,b; Secrest et al. 2021; Siewert et al. 2021; Singal 2021a,b) 
%\cite{4,5,6,7,8,9,10,11,Si21a,Si21b,Si21c} 
that the AGNs, which presumably are the best representatives of the average distribution of matter in the universe at large redshifts ($z\stackrel{>}{_{\sim}} 1$), do not seem to share the CMBR reference frame at much larger redshifts ($z\stackrel{>}{_{\sim}} 10^3$). Therefore it is necessary that we keep an open mind and try to determine the dipole that might represent the peculiar motion of the observer, moving as a part of the Solar system with respect to the average matter universe, by using as many  independent datasets, employing different methods, as possible.

In essence, the advantages of the present technique stem from (i) the Hubble diagram method for determining peculiar motion of the observer does not depend upon the exact nature of the $m-z$ relation or the underlying physics of the objects used in the sample. Nor does it depend upon the particular cosmological model to fit the $m-z$ relation. All that is required is a simple empirical relation, that could be discerned in the $m-z$ plot. (ii) The completeness of the survey, where all sources above a flux density limit may be included in the basic surveys from which the sample may be derived, may not be a prerequisite here. (iii) The sample may not be having a uniform coverage of the sky; a piece-wise coverage of the sky in different directions could suffice, as long as the sources cover various different directions in the sky so as to sample different projected components of the peculiar velocity. Therefore gaps in sky coverage or clustering of sources may not affect the results, as long as no other bias in $m-z$ plot, apart from the effects of the peculiar motion itself, gets introduced. 

\begin{figure*}
\includegraphics[width=\linewidth]{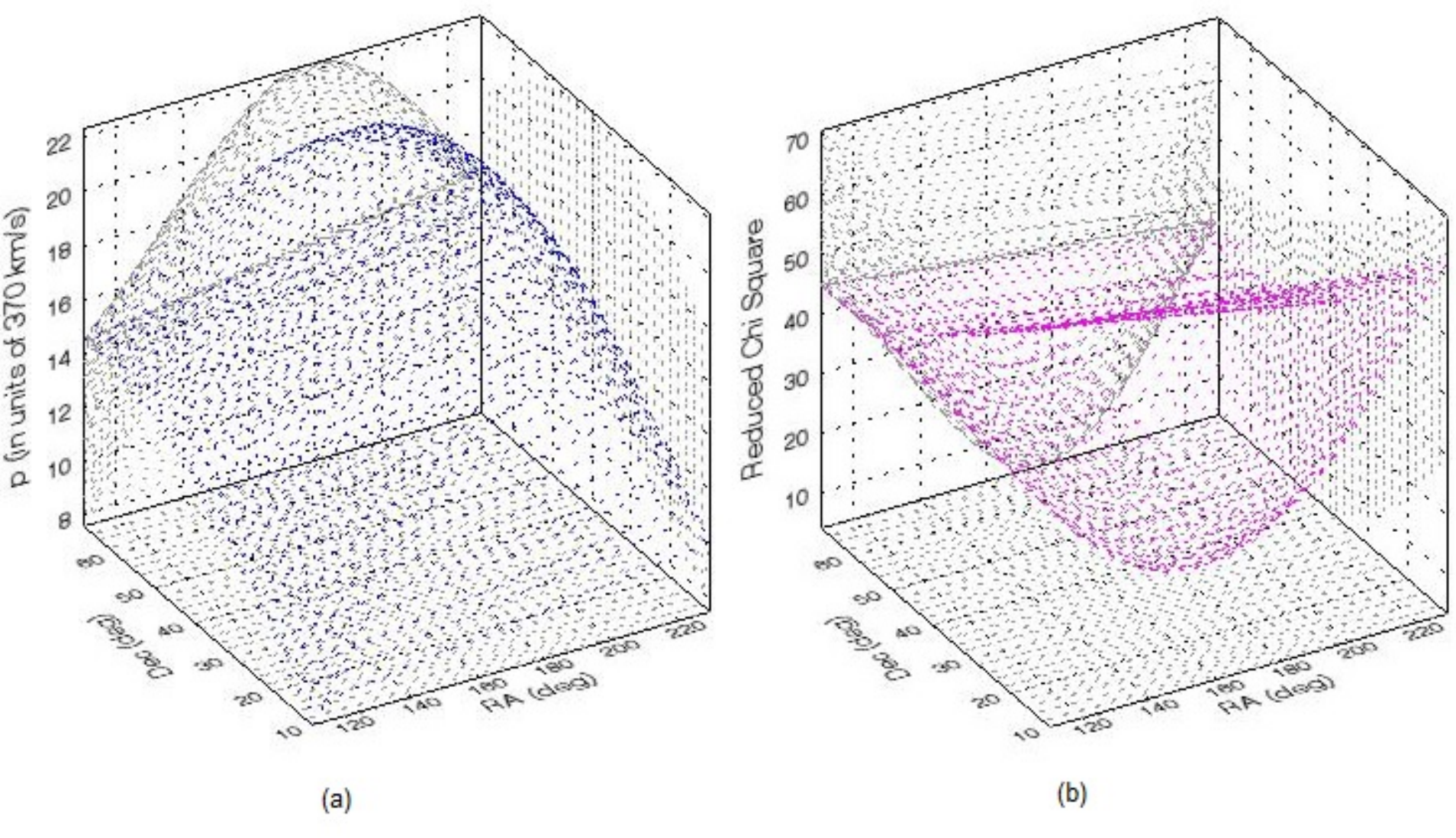}
\caption{A 3-d plot (a) of the COSFIT routine result (in blue colour) for the trial dipole directions across the sky, showing a peak at RA$=180^{\circ}$, and Dec$=42^{\circ}$ (b) of the reduced chi-square ($\chi^2_\nu$) values (in violet colour), from the cosfit routine for various trial directions for the dipole. The horizontal plane shows the direction in sky as RA and Dec in degrees. The position of the extremum in each case is determined easily from the 2-d projections, shown in light grey. 
The minimum chi-square value of 3.7, somewhat higher than the ideal value of unity, occurs at RA$=177^{\circ}$, and Dec$=42^{\circ}$, 
We infer the optimum direction of the observer's peculiar velocity to be RA$=179^{\circ}\pm 25^{\circ}$, and Dec$=42^{\circ}\pm 25^{\circ}$, which lies within $2\sigma$ of the CMBR Dipole direction, however has a rather large amplitude (higher by a factor of $22\pm 5$).} 
\end{figure*}

%---------------------------------------------------

This technique has been successfully applied to determine the peculiar motion of the solar system, from the magnitude-redshift Hubble diagram for SNe~Ia, which are one of the best standard candles known, with a very tight $m_{\rm B}-z$ relation, where $m_{\rm B}$ denotes the observed peak in blue magnitude and $z$ the measured redshift of each SN~Ia. This technique could yield statistically significant results from a much smaller number of ($\stackrel{<}{_{\sim}} 10^3$) SNe Ia (Singal 2021c). 

In fact, this technique could as such be applied to a combination of data from a heterogeneous set of various sub-samples; all that is required is that no systematic errors have observationally entered in the redshift and magnitude estimates of individual sources depending upon the direction in sky.  A very non-uniform sky coverage, as seen in Fig.~3, would have made it an almost meaningless exercise to try to get the peculiar motion from number counts or sky brightness, while from the Hubble diagram, as we shall show, we should be able to get values of the peculiar motion, even if with somewhat large errors.

On the negative side, the absence of a clear cut simple $m-z$ relation, for instance a straight line fit of the type seen in Fig.~1, could make it difficult to figure out the differential effects of the peculiar motion for sources in opposite hemisphere $\Sigma_1$ and $\Sigma_2$. Or even a large spread in the magnitudes at any given redshift would increase the statistical uncertainties in the estimated peculiar velocity, both in its direction and amplitude. Another difficulty could arise from the fact that in the $m-z$ plot, the deviations providing a measure of the observer's peculiar motion become more significant as one goes to lower redshifts ($z < 0.05$) where, however, the presence of local bulk motions, that could be correlated with the solar peculiar motion, might introduce some erroneous influence on the results. In order to contain that, one may need to exercise extra caution about the sources in the sample at such low redshifts ($z \stackrel{<}{_{\sim}}  0.05$).
%-----------------------------------------------------------------
\section{Determining the direction and amplitude of the peculiar motion}
%--------------------------------------------
%--------------------------------------------
\begin{figure*}
\includegraphics[width=\linewidth]{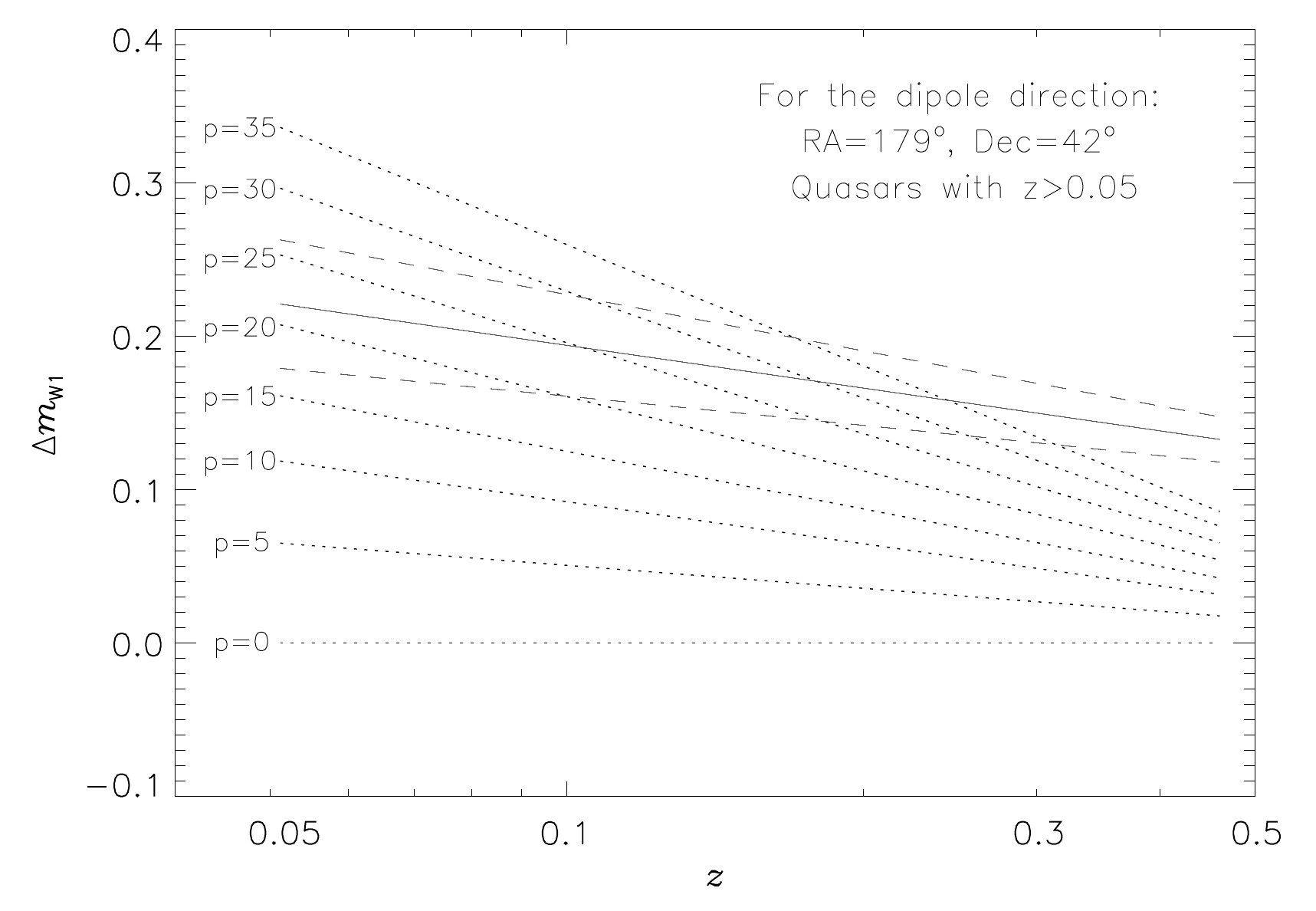}
\caption{For observer's peculiar velocity direction along RA=179$^\circ$, Dec=42$^\circ$, the differences expected in magnitudes, $\Delta m_{\rm w1}$, as a function of redshift ($z$), for  quasars with redshift limits, $0.05< z< 4$, from the two hemispheres, $\Sigma_1$ and $\Sigma_2$, plotted as dotted lines, for different peculiar velocity values ($p=0$ to 35). 
The unbroken line shows a fit to the actual observed difference between the average magnitudes of sources from the two hemispheres, at different redshifts, while the dashed lines above and below the unbroken line represent the $1 \sigma$ uncertainties in the fit.}
\end{figure*}
%--------------------------------------------
%--------------------------------------------
\begin{figure*}
\includegraphics[width=\linewidth]{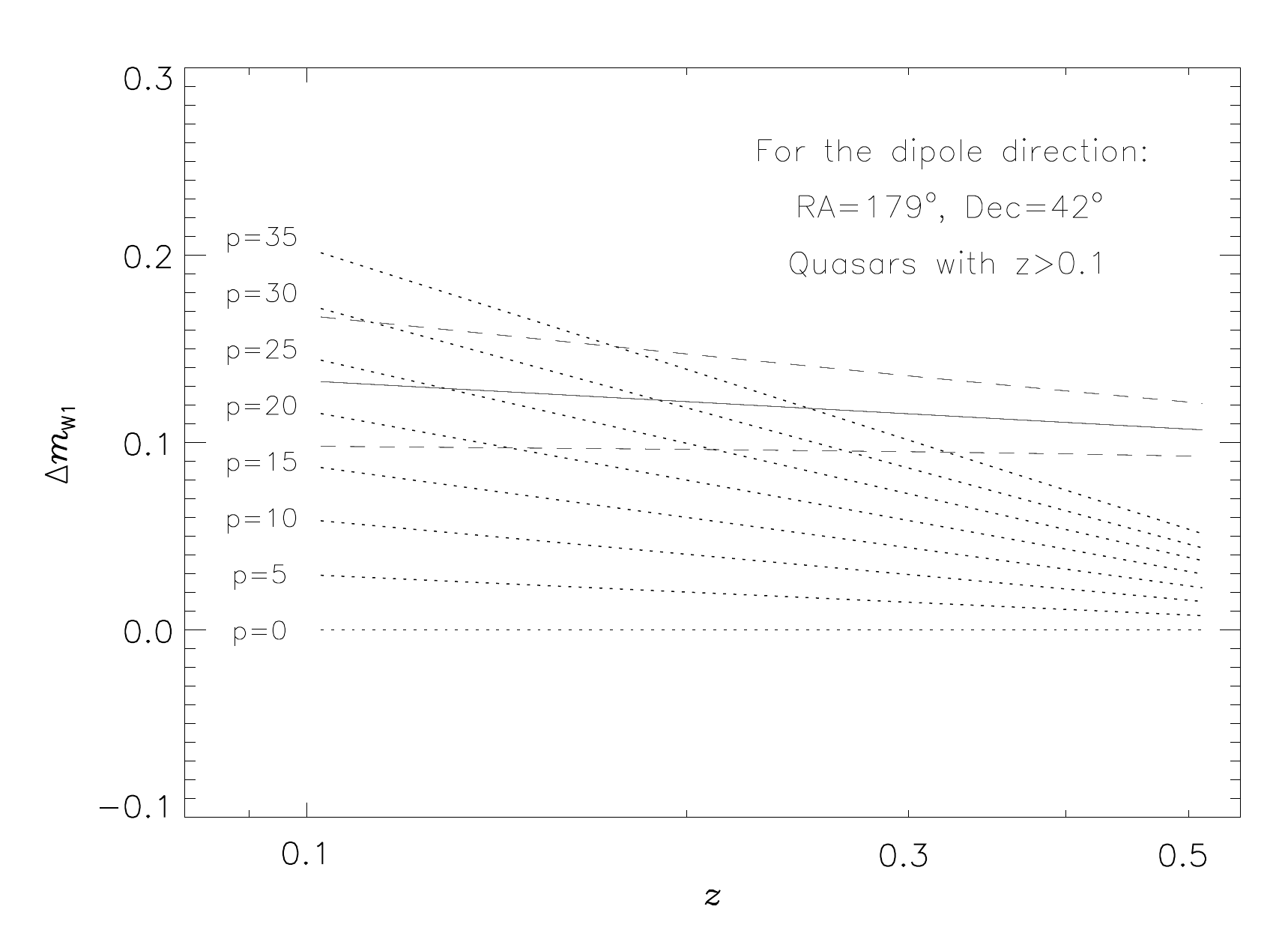}
\caption{For observer's peculiar velocity direction along RA=179$^\circ$, Dec=42$^\circ$, the expected differences in magnitude, $\Delta m_{\rm w1}$, as a function of redshift ($z$), for quasars with redshift limits, $0.1< z< 4$, from the two hemispheres, $\Sigma_1$ and $\Sigma_2$, plotted as dotted lines, for different peculiar velocity values ($p=0$ to 35). 
The unbroken line shows a fit to the actual observed difference between the average magnitudes of sources from the two hemispheres, at different redshifts, while the dashed lines above and below the unbroken line represent the $1 \sigma$ uncertainties in the fit.}
\end{figure*}
%--------------------------------------------

Since the quasar dipole might not lie in the same direction as that of any previously known dipoles, including that of the CMBR, we want to determine the quasar dipole in a manner that is unbiased toward any particular direction. To achieve that, we have employed the `brute force' method (Singal 2019b), where we first divided the sky into $10^\circ \times 10^\circ$ pixels with minimal overlap, creating a grid of $422$ cells covering the whole sky. 
Then taking the dipole direction to be the centre of each of these 422 cells, in turn, we computed the dipole magnitude $p$ (along with standard error) that provided the best fit to the $m_{\rm w1}-z$ data for the quasars in our sample. Actually this yields only a projection of the peculiar velocity in the direction of the pixel being tried. From all 422 $p$ values, a 
broad plateau showing a maxima towards certain range of directions, not too far from the CMBR dipole direction, was though discernible, however, due to fluctuations in individual $p$ values it was not possible to zero down on a single unique peak for the dipole direction. Since with respect to the actual dipole direction there should be a $\cos\psi$ dependence of the $p$ values  determined for other grid points, we made a 3-d $\cos\psi$ fit for each of the $n=422$ positions for the remaining $n-1$ $p$ values, and also computed the chi-square value for each of these $n$ fits. 

The 3-d COSFIT routine resulted in a clear unique peak, with an accompanying minimum chi-square value in the near vicinity of the peak. The peak occurred at RA$=180^{\circ}$, Dec$=42^{\circ}$, with a height, $p=22$. Figure 5 shows the 3-d COSFIT plot of the peak as well as a clear minimum in the reduced $\chi ^2$ value at RA$=178^{\circ}$, Dec$=42^{\circ}$.  In order to make sure, we also tried finer grids with $5^\circ \times 5^\circ$ bins with $1668$ cells and even a grid with $2^\circ \times 2^\circ$ bins with $10360$ cells, but it made no perceptible difference in our results. 
Figure 5 shows the outputs of our COSFIT routine.

%---------------------------------------------------
To test our COSFIT procedure, we made simulations, with random positions (RA and Dec) in sky allotted to quasars in our sample and then a mock dipole in sky was superimposed to calculate $z$ and $m_{\rm w1}$  for each source according to Eqs.~({\ref{eq:80.1}) and ({\ref{eq:80.2}). Then on this mock catalogue of quasars, our procedure was applied to recover the dipole and compared with the input dipole in that simulation. This not only validated our method but it also provided us an estimate of errors from 1500 independent simulations, in three sets of 500 each, so as to be sure that results from simulations, including error estimates, are consistent across different sets of  simulations.

Accordingly, for the direction of the peculiar velocity we arrive at RA$=179^{\circ}$, Dec$=42^{\circ}$ from the minimum value of the Chi-square fit as well as from the peak in dipole magnitude, with both almost coinciding. Estimated errors in the dipole are $\Delta$RA$=\pm 25^{\circ}$ and $\Delta$Dec$= \pm25^{\circ}$. This value for the derived quasar dipole direction agrees, within the $2\sigma$ uncertainty, with the CMBR dipole direction, RA$=168^{\circ}$, Dec$=-7^{\circ}$. However the magnitude of the quasar dipole corresponds to with $p=22\pm 5$, or a solar speed $8.1\pm 1.8 \times 10^3$ km s$^{-1}$, a $4.5\sigma$ result. 

Figure 6 shows the differences expected in magnitudes, $\Delta m_{\rm w1}$, as a function of redshift ($z$), between sources from the two hemispheres, $\Sigma_1$ and $\Sigma_2$, now placed with respect to the pole along observer's peculiar velocity direction, RA=179$^\circ$, Dec=42$^\circ$, plotted as dotted lines, for different peculiar velocity values ($p=0$ to 35). 
The unbroken line shows a fit to the actual observed difference between the average magnitudes of sources from the two hemispheres, at different redshifts , while the dashed lines above and below the unbroken line represent the $1 \sigma$ uncertainties in the fit. We get $p=22\pm 5$ for the peculiar velocity.

The large value of inferred $p$ might raise some doubts whether it is due to some contamination in the data, e.g., due to low redshift quasars ($z \sim 0.05$) lying among local superclusters, and their local bulk motion, that might be correlated with the Solar peculiar motion, adding to the inferred $p$ value. In order to either verify or rule it out we raised the lower redshift limit to 0.1 in our sample, which left 114495 quasars in our sample for which we redetermined the value of $p$. Figure 7 shows the result where we get $p=23\pm 6$ for the peculiar velocity. It shows that the large value of $p$ is not due to some suspected contaminations from nearby quasars at $z\sim 0.05$, and that the derived value of $p$ is not much affected even when we restrict the quasar redshifts to be $>0.1$. 
%Why it is higher than all previously published values, obtained by other methods, is a question which may need to be investigated separately. 

Since the method does not depend upon the number of sources in the sample, apart from that the errors may increase for a smaller number of sources, we split our sample into two, by picking the odd and even serial number sources in the original sample and separately combining them to get two sub-samples and then following our above procedure, determined the peculiar motion from the two sub-samples independently. The results that we thus got for the dipole values were $p=22\pm 6.5$ along RA$=185^{\circ}$, Dec$=46^{\circ}$ and  $p=22\pm 6$ along RA$=172^{\circ}$, Dec$=38^{\circ}$. These are consistent with the value, $p=22\pm 5$ along RA$=179^{\circ}$, Dec$=42^{\circ}$ derived from the whole sample, with no unexpected variations in the results from the two sub-samples. This supports the view that the shift in the Hubble diagram between the two set of sources belonging to opposite hemispheres is a genuine effect, presumed to be a result of the observer's peculiar motion, and not arising from some skew distribution among some small number of  sources in either dataset.  

%------------------------------------------------------------------------
It may be pointed out that a search for a redshift anisotropy in a homogeneously selected sample of $ 103245$ SDSS quasars (Singal 2019b) had earlier yielded a significant redshift dipole along  the same direction as the CMBR dipole, which, however, when interpreted due to a Solar peculiar motion, gave for the peculiar velocity a direction opposite to the CMBR dipole, with an amplitude $p\approx 6.5$. On the other hand, the Hubble $m_{\rm w1}-z$ plot shows an infrared magnitude ($m_{\rm w1}$) dipole with brighter sources in a direction opposite to the CMBR dipole (Fig.~2), which when interpreted as due to the Solar peculiar motion,  gives a peculiar velocity along the CMBR dipole with $p=22$, about 3.5 times larger than that estimated from the SDSS quasar sample. 
%Moreover, the directions derived for the peculiar velocities seem to be lying along opposite points in sky. 
This could be puzzling as the two basic samples (WISE and SDSS quasars) might have major overlap. A detailed comparison of the two chosen samples (115610 WISE quasars vs. 103245 SDSS quasars) has shown, however, that the overlap in these two particular samples is rather small ($<10\%$), with 
%only 9333 out of 115610 sources in our CatWISE sample are common with the SDSS quasars, 
a very large majority of sources in the two samples ($>90\%$) being independent. In fact the   $m_{\rm w1}$ dipole information in the present technique comes mostly from the data at low redshifts (($z \stackrel{<}{_{\sim}} 0.5$) (Figs.~6 and 7), where the overlap in the two samples is minuscule ($<0.5$ percent).
%; 91 out of 19051).
%For sources z<0.6, it is 139 out of 24349 =0.57%, 
%For sources z<0.5, it is 91 out of 19051 =0.48%

In any case. even with a minimal overlap, one should expect the two dipole values to be consistent, provided of course, that both are resulting from the same basic phenomenon, viz. Solar peculiar motion in this case. With many more spectroscopic redshifts for quasars over substantially larger regions of the sky that might get provided by the Euclid satellite (Amendola et al. 2016) or from the Legacy Survey of Space and Time (LSST) made with the Rubin Observatory (Ivezić et al. 2019), it may be possible to resolve this dichotomy of widely separated dipole magnitudes. However, if the genesis of the dipoles is a result of some other phenomenon, depending upon their sources of origin, for instance intrinsic anisotropies in the sky distributions of their infrared magnitudes or redshifts, the two dipoles could still remain different from each other. 
%in magnitude, even if the directions are the same, depending upon their basic causes.
After all, while in the present investigation we looked for a dipole in infrared magnitudes ($m_{\rm w1}$), in the case of SDSS quasars it was the redshift ($z$) dipole that was being investigated. 
%The two dipoles individually ($m_{\rm w1}$ magnitudes and the redshift values) seem to point in the same direction as the CMBR dipole. Of course if we interpret these two dipoles in terms of Solar system peculiar velocity, then the two seem to point in opposite directions.  
It might be that these partly or wholly are not mere kinematic dipoles, representing Solar peculiar motion, and instead are some intrinsically different (as yet unknown!) dipoles. 

That the kinematic origin might not be the whole truth for these cosmic dipoles became evident from the very first successful dipole determination from number count of radio galaxies (Singal 2011) that showed a much larger value for the dipole (by a factor of ~4) as compared to the CMBR dipole, contrary to what expected from conventional wisdom. Because the directions seemed to match, showed that it may not be due to some systematics, otherwise even the directions should have turned out very different. It indicated that these might represent some genuine, intrinsically different, dipoles and not just the kinematic dipole. Later observational results over the last decade 
have confirmed these results whenever a dipole from the same survey, e.g. the NVSS, was determined by independent workers (Rubart \& Schwarz 2013; Tiwari et al. 2015; Colin et al. 2017; Bengaly et al. 2018; Singal 2019a). At the same time, results from different surveys have yielded dipoles with an order of magnitude differences in the amplitudes (Bengaly et al. 2018; Singal 2019a,b; Siewert et al. 2021; Secrest et al. 2021; Singal 2021a,b,c). For instance, the number count method yielded a rather large value of the peculiar velocity ($p \sim 10-15$), at a significantly high sigma level, in the TGSS survey (Bengaly et al. 2018; Singal 2019a; Siewert et al. 2021).
These puzzling results over the last decade have already cast serious doubts on the kinematic interpretation and more so, on the CP itself. It may be mentioned that because of the much higher value of the dipole from the TGSS data, which at 150 MHz is at much lower frequency as compared to  say, the NVSS data at 1400 MHz, a frequency dependence of the radio dipole strength, $p(\nu) \propto \nu ^m$ with $m \sim -0.5$, has been suggested (Siewert et al. 2021).
But this inference may need to be treated with caution as it gets drawn basically from a single high value point, viz. TGSS dipole amplitude (Siewert et al. 2021, Fig. 9).

Quasars provide a powerful probe of structure formation in the universe. Such structures are likely to be found at the highest peaks of the density field, and would thus be highly biased tracers (Kaiser 1984; Bardeen et al. 1986; Djorgovski 1999; Shen et al. 2009; Retana-Montenegro \&  Röttgering 2017; Alam et al. 2021) of the overall mass distribution in the Universe. Nevertheless, as per the cosmological principle, the sky distribution of even the highest peaks of the density field should be more or less isotropic as well. Thus even if quasars may be highly biased tracers of the matter distribution in the universe, these should still serve as good probes to test the  cosmological principle, our main aim here. Of course, at lower redshifts, this bias could become important due to local clustering. However, in our investigation, we have tried to minimize its effect, if any, by restricting the lower limit of redshifts, z>0.05, in our sample of quasars. Moreover, a change in the lower limit to z>0.1 hardly made any difference to the results (cf. Figs. 6 and 7), ruling out any significant role of such a bias in our results.

A sky survey at multiple frequencies, carried out with the Square Kilometre Array (SKA) 
could be used for an independent investigation of the controversial dipole anisotropy with a much superior sensitivity and thereby settle the question of the CP, hopefully, in a more decisive manner. In fact it was argued by Crawford (2009) that it is not possible to detect a radio dipole at more than $1\sigma$ level (for a presumed radio dipole amplitude equal to the CMBR dipole, i.e. $p=1$), in a survey like the NVSS as the latter does not have sufficient number of sources. The conclusion drawn instead was that in order to make a positive detection of a radio dipole we may have to wait for the SKA data, going to sub-$\mu$Jy flux-density levels, and thus having much larger number of sources, to get a better signal to noise. 
In spite of this deterrent prediction, a radio dipole at statistically significant level was  detected from the NVSS data itself (Singal 2011) which became possible only because the radio dipole amplitude turned out to be actually a factor of $\sim 4$ larger than the CMBR dipole. Although  SKA might provide number counts at sub-$\mu$Jy levels (Schwarz et al. 2015; Bengaly et al. 2019),
however, one may need to be wary of possible caveats in using radio source number counts at 
these flux-density levels as one might start seeing at sub-mJy levels, in addition to the powerful distant radio sources, a substantially increasing fraction of very different populations of radio sources, e.g., nearby normal galaxies, star burst galaxies and even galactic sources (Windhorst 2003; Padovani 2011; Luchsinger 2015). Instead what would be important is to get surveys at different frequencies from SKA, at a few mJy levels or above where the radio source population may comprise mostly powerful radio galaxies and quasars, and then investigate the dipoles to see how genuine is the difference in dipoles from surveys at widely separated frequency bands like that seen in the NVSS and TGSS dipoles. 
The multiple frequency flux-density measurements should also allow for the chosen sample a direct estimate of an optimum value of the spectral index, that enters in the expression for the dipole amplitude. 
The question of CP could be resolved in a positive manner if it is found that  multiple-frequency radio source surveys yield dipoles consistent with the CMBR dipole, though it would still remain to be explained why the presently determined dipoles are not consistent with that. However, in case it does turn out that the SKA surveys with better statistical accuracies also yield dipole magnitudes which are significantly different from the CMBR dipole, even though might be pointing along the CMBR dipole direction, as is seen in the presently determined dipoles, then it will certainly be a big problem for the CP, and consequently all conventional cosmological models, including the ones dealing with the question of dark energy, could be in serious jeopardy. 

%---------------------------------------------------------------------
\section{Implications for the Cosmological Principle}
Peculiar motion of the Solar system can provide a direct test of the CP, according to which the various cosmic reference frames should be 
coincident with the reference frame defined by the CMBR, with no relative motion with respect to it. However, as was first pointed out in 
Singal (2011), the reference frame defined by the NVSS radio data does not coincide with the CMBR reference frame. Subsequent 
investigations have repeatedly shown that not only various cosmic reference frames seem to have relative motion with respect 
to the CMBR reference frame, they do not seem to coincide among themselves (Rubart \& Schwarz 2013; Tiwari et al. 2015; Colin et al. 2017; Bengaly et al. 2018; Singal 2019a,b; Siewert et al. 2021; Secrest et al. 2021; Singal 2021a,b,c). Our present reference frame determined from the Hubble diagram of quasars, in fact, seems to differ the most. All these are inconsistent with the CP, in any case.
%--------------------------------------------
\begin{figure}
\includegraphics[width=\columnwidth]{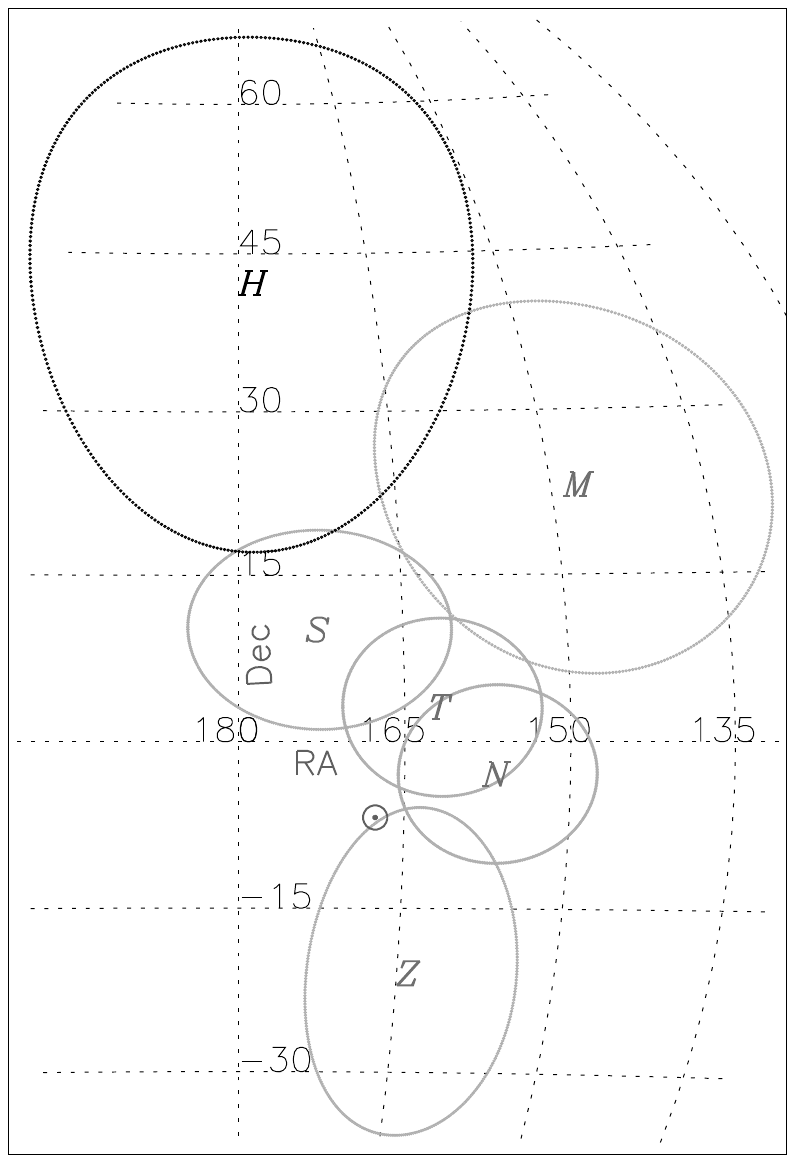}
\caption{A small 
portion of the sky, in~the Hammer--Aitoff equal-area projection, plotted in equatorial coordinates RA and Dec, showing the position of~the pole determined from the Hubble diagram of our quasar sample, indicated by $H$, along with the error ellipse. Also shown on the map are the other pole positions for various dipoles along with their error ellipses,  $N$ (NVSS), $T$ (TGSS), $Z$ (DR12Q),  $M$ (MIRAGN), $S$ (SNe Ia). The~CMBR pole, at RA$=168^{\circ}$, Dec$=-7^{\circ}$, indicated by $\odot$, has negligible errors.}
\end{figure}
%--------------------------------------------

In order to compare directions of the dipoles determined from different datasets, we show  in Fig. 8 the relative positions of the 
estimated directions of various dipoles in a small, relevant portion of the sky. The position of~the pole determined from the Hubble diagram 
of our quasar sample, indicated by $H$, is shown along with the error ellipse. Also shown are the pole positions for other dipoles, 
along with their error ellipses: $N$ (NVSS) (Singal 2011), $T$ (TGSS) (Singal 2019a), $Z$ (DR12Q) (Singal 2019b),  $M$ (MIRAGN) 
(Singal 2021a,b}, $S$ (SNe Ia) (Singal 2021c). The~CMBR pole, at RA$=168^{\circ}$, Dec$=-7^{\circ}$, indicated by $\odot$, has 
negligible errors (Aghanim et al. 2020). It seems that the poles of the dipole direction from the Hubble diagrams of quasars and SNeIa as well as from number counts of mid-infrared 
quasars lie within $2\sigma$ of the CMBR pole, but each of the other three dipoles lies within $\sim 1\sigma$ of the CMBR pole. From that 
we can surmise that the various dipoles, including the CMBR~dipole, are all pointing along the same direction.
Nevertheless, as we mentioned earlier, all these other dipoles have much larger amplitudes than the CMBR dipole, with almost an order 
of magnitude spread, even though various dipole directions in the sky may be lying parallel to each other.
From various dipoles we cannot arrive at a single coherent picture of the solar peculiar velocity, which,   
defined as a motion relative to the local comoving coordinates  and from the CP, a motion with respect to an average universe, 
should after all not depend upon the exact method used for its determination. 

Since a genuine solar peculiar velocity cannot vary from one dataset to the other, an order of magnitude, statistically significant, incongruous dipole amplitudes, might imply that we may instead have to look for some other cause for the genesis of these dipole, including that of the CMBR, the latter generally assumed by the conventional wisdom to be of kinematic origin owing to the Solar peculiar motion. 
The observed fact that different dipoles, resulting from different, independent datasets, obtained with independent instruments and techniques in different wavebands by different independent groups, happen to be pointing along the same direction in sky, shows that these dipoles are not results of some systematics in individual datasets, otherwise they would have been pointing in random, independent directions in sky. This particular direction in sky must have some peculiarity and is in some respects a preferred direction, a sort of an ``axis'' of the universe. In any case, so many independent dipole vectors pointing along the same particular direction could imply an inherent  anisotropy, which, in turn, would be against the CP, the most basic tenet of the modern cosmology.

In literature, there are many alternative models put forward for the dipoles, e.g., the Hubble parameter being higher in the direction of the CMBR dipole at higher redshifts ($z\sim 1$) has been suggested (Krishnan et al. 2021a,b), or 
there is a model suggested for a ``dark flow'' dipole at the position of galaxy clusters, found in filtered maps of the CMBR temperature anisotropies, implying the existence of a primordial CMBR dipole of non-kinematic origin (Kashlinsky et al. 2008,09,10; Atrio-Barandela et al. 2015).
% or  some dynamical contributions to the CMBR dipole, resulting possibly from an SU(2) gauge principle~\cite{Sz08,Lu09,Ho13}. 
Alignments among some other dipoles have also been pointed out. Based on a study of quasar absorption systems, a dipole in the spatial variation of the fine-structure constant, $\alpha = e^2/\hbar c$ has been reported (Webb et al. 2011; Berengut et al. 2011; King et al. 2012;  Berengut, Kava \& Flambaum 2012) and these spatial variations of $\alpha$ could be constrained using clusters of galaxies (Galli 2013; Martino et al. 2016). It has been claimed that the fine structure constant cosmic dipole is aligned with the corresponding dark energy dipole within 1$\sigma$ uncertainties (Mariano \& Perivolaropoulos 2012,13).
Further, the odd multipoles in the large scale anisotropies of the CMB temperature, a~parity asymmetry, show a preference to~be strongly aligned with the CMBR dipole at significant levels (Naselsky et al. 2012; Zhao 2014; Cheng et al. 2016). 
All such alignments are inconsistent with the CP and the ensuing standard model. 

%--------------------------------------------
\section{Conclusions}
From the Hubble diagram of quasars, peculiar motion of the Solar system is derived, which turns out to be the largest value ever found, $\sim 22$ times larger than that inferred from the CMBR dipole, though the direction lies within $2\sigma$ of the CMBR dipole. It seems that the dipoles determined from the AGNs, or even from SNe Ia, somehow have much larger amplitudes than the CMBR dipole, with  almost an order of magnitude spread, concordant neither with each other nor with the CMBR, though their directions coincide within statistical uncertainties. It shows that they are not randomly oriented and that the various dipoles, including the CMBR~dipole, all pointing along the same direction, 
suggest a preferred direction in the Universe, raising thereby uncomfortable questions about the CP, the basis of the standard model in modern cosmology.
%%%%%%%%%%%%%%%%%%%%%%%%%%%%%%%%%%%%%%%%%%
\section*{Declarations}
The author has no conflicts of interest/competing interests to declare that are relevant to the content of this article. No funds, grants, or other support of any kind was received from anywhere for this research.
%%%%%%%%%%%%%%%%%%%%%%%%%%%%%%%%%%%%%%%%%%
%--------------------------------------------
\section*{Data Availability}
The data underlying this article is freely available at VizieR Astronomical Server in the public domain (http://vizier.u-strasbg.fr/viz-bin/VizieR) and can be downloaded by selecting the catalog: J/ApJS/221/12/table1. 
%--------------------------------------------
%--------------------------------------------


\begin{thebibliography}{20}
\bibitem{16.1} Aghanim N., Akrami T., Arroja F. et al., 2020, A\&A, 641, A1
\bibitem{Al21} Alam S., Ross N. P., Eftekharzadeh S., Peacock J. A., Comparat J.,
 Myers A. D., Ross A. J., 2021, MNRAS, 504, 857
\bibitem{Am16} Amendola L., Appleby S., Avgoustidis A. et al., 2016, arXiv:1606.00180
\bibitem{At15} Atrio-Barandela F., Kashlinsky A., Ebeling H., Fixsen D. J., Kocevski D., 2015, ApJ, 810, 143
\bibitem{Ba86} Bardeen J., Bond J. R., Kaiser N., Szalay A. S., 1986, ApJ, 304, 15
\bibitem{12} Bengaly C. A. P., Maartens R., Santos M. G., 2018, J. Cosm. Astropart. Phys., 4, 31
\bibitem{12a} Bengaly C. A. P.,  Siewert T. M., Schwarz D. J., Maartens R., 2019, MNRAS, 486, 1350
\bibitem{Be11} Berengut, J. C., Flambaum, V. V., King, J. A., Curran, S. J., Webb, J. K., 2011, Phys. Rev. D,  {83}, 123506 
\bibitem{Be12} Berengut J. C., Kava E. M., Flambaum V. V., 2012, A\&A, {542}, A118 
%\bibitem{Bo15} Bourke T. L., Braun R., Fender R. et al., 2015, Advancing Astrophysics with the Square Kilometre Array in Proc. Sci., 174 
\bibitem{Ch16} Cheng, C., Zhao, W., Huang, Q.-G., Santos, L., 2016, Phys. Lett. B, {757}, 445 
\bibitem{} Colin J., Mohayaee R., Rameez M., Sarkar S., 2017, MNRAS, 471, 1045
%\bibitem{16} Colin J., Mohayaee R.. Rameez M., Sarkar S., 2019, A\&A,  631, L13
\bibitem{35} Condon J. J., Cotton W. D., Greisen E. W., Yin Q. F., Perley R. A., Taylor G. B., Broderick J. J., 1998, AJ, 115, 1693
%\bibitem{21a} Condon J. J., 1992, ARA\&A, {30}, 575
\bibitem{} Crawford F., 2009, ApJ, 692, 887
\bibitem{21} Davis T. M., Hui L., Frieman J. A. et al., 2011, ApJ, 741, 67 
\bibitem{Dj99} Djorgovski S. G., 1999, in The Hy-Redshift Universe: Galaxy Formation and Evolution at High Redshift, eds. A.J. Bunker  W.J.M. van Breugel, \aspc 
\bibitem{Ga13} Galli, S., 2013, Phys. Rev. D, {87}, 123516 
\bibitem{32} Hinshaw G., Weiland J. L., Hill R. S. et al., 2009, ApJS, 180, 225
\bibitem{11} Intema H. T., Jagannathan P., Mooley K. P., Frail D. A., 2017, A\&A, 598, A78, 
\bibitem{Iv19} Ivezić Ž., Kahn S. M., Tyson J. A. et al., 2019, ApJ, 873, 111
\bibitem{Kai84} Kaiser N., 1984, ApJ, 284, L9
\bibitem{Ka08} Kashlinsky, A., Atrio-Barandela, F., Kocevski, D., Ebeling, H., 2008, ApJ, {686}, 498
\bibitem{Ka09} Kashlinsky, A., Atrio-Barandela, F., Kocevski, D., Ebeling, H., 2009, ApJ, {691}, 1479
\bibitem{Ka10} Kashlinsky, A., Atrio-Barandela, F., Ebeling, H., Edge, A., Kocevski, D., 2010, ApJ, {712}, L81  
\bibitem{Ki12} King, J. A., Webb, J. K., Murphy, M. T., Flambaum, V. V., Carswell, R. F., Bainbridge, M. B., Wilczynska, M. R., Koch, F. E., 2012, MNRAS, {422}, 3370
\bibitem{Kr21a} Krishnan, C., Mohayaee, R., \'O Colg\'ain, E., Sheikh-Jabbari, M. M., Yin, L., 2021a, arXiv:2105.09790  
\bibitem{Kr21b} Krishnan, C., Mohayaee, R., \'O Colg\'ain, E., Sheikh-Jabbari, M. M., Yin, L., 2021b, arXiv:2106.02532
\bibitem{31} Lineweaver C. H., Tenorio L., Smoot G. F., Keegstra P., Banday A. J., Lubin P., 1996, ApJ, 470 38
\bibitem{23} Luchsinger K. M., Lacy M., Jones K. M. et al.,2015, AJ, 150, 87
\bibitem{19} Mainzer, A., Bauer, J., Cutri, R. M. ~et~al., 2014, ApJ, {792}, 30
\bibitem{Ma12} Mariano, A., Perivolaropoulos, L., 2012, Phys. Rev. D, {86}, 083517
\bibitem{Ma13} Mariano, A., Perivolaropoulos, L., 2013, Phys. Rev. D, {87}, 043511 
\bibitem{Mar16} Martino, I. De, Martins,  C. J. A. P., Ebeling, H., Kocevski, D., 2016, Phys. Rev. D, {94}, 083008
\bibitem{Na12} Naselsky, P., Zhao, W., Kim, J., Chen, S., 2012, ApJ, {749}, 31
\bibitem{24} Padovani, 2011, MNRAS, {411}, 1547
\bibitem{Re17} Retana-Montenegro E.,  Röttgering H. J. A., 2017, A\&A, 600, A97 
\bibitem{33} Rubart M., Schwarz D. J., 2013, A\&A, 555, A117
\bibitem{Sc15} Schwarz D. J., Bacon D., Chen S. et al. 2015, in Proc. Sci. (AASKA14), eds. Bourke T. L., Braun R., Fender R. et al., 14, 1461 (arXiv:1501.03820)
\bibitem{} Secrest N. J., Dudik R. P., Dorland B. N., Zacharias N., Makarov V., Fey A., Frouard J., Finch, C., 
%Identification of 1.4 million active galactic nuclei in the mid-infrared using WISE Data. 
2015, \apjs, 221, 12 
\bibitem{9} Secrest N. J., Hausegger S. V., Rameez M., Mohayaee R., Sarkar S., Colin J., 2021,  ApJ, 908, L51 
\bibitem{Sh09} Shen Y. et al., 2009, ApJ, 697, 1656
\bibitem{Sie21} Siewert T. M., Rubart M. S., Schwarz D. J.,  2021, A\&A, 653, A9
\bibitem{4} Singal A. K., 2011, ApJ, 742, L23 
\bibitem{10b} Singal A. K., 2019a, Phys. Rev. D, 100, 063501
\bibitem{10a} Singal A. K., 2019b, MNRAS, 488, L104 
\bibitem{si21a} Singal, A. K., {2021a}, 
%Solar system peculiar motion from mid infra red AGNs and its  cosmological implications. 
%Proc. 1st Electronic Conference on Universe 2021 (ECU2021), 
Phys. Sci. Forum, 2, Issue 1, 54
\bibitem{si21b} Singal A. K., 2021b, Universe 7, 107
%, doi:10.3390/universe7040107
\bibitem{si21c} Singal A. K., 2021c, arXiv:2106.11968
\bibitem{14} Swarup G., Ananthakrishnan S., Kapahi V. K., Rao A. P., Subrahmanya C. R., Kulkarni V. K. 1991, Current Science, 60, 95
\bibitem{34} Tiwari P., Kothari R., Naskar A., Nadkarni-Ghosh S., Jain P., 2015, Astropart. Phys., 61, 1
\bibitem{Web11} Webb, J. K., King, J. A., Murphy, M. T., Flambaum, V. V., Carswell, R. F., Bainbridge,  M. B., 2011, Phys. Rev. Lett., {107}, 191101
\bibitem{22} Windhorst R. A., 2003, \nar, 47, 357
\bibitem{18} Wright, E. L., Eisenhardt, P. R. M., Mainzer, A. K. et al., 2010 AJ, {140}, 1868
\bibitem{Zh14} Zhao, W., 2014, Phys. Rev. D, {89}, 023010 
\end{thebibliography}
\end{document}